\documentstyle[aps,preprint,epsfig,epsf]{revtex}
\begin {document}
\draft
\title{Study of pseudogap in underdoped cuprate}
\author{Ping Lou}
\address{Department of Physics, Anhui University, Hefei 230039, People's Republic of China} 
\author{Hang-sheng Wu}
\address{Department of Physics, University of Science and Technology of China,
                     Hefei 230026, People's Republic of China}
\maketitle
\begin{abstract}
A mean field SDW analysis of pseudogap in the underdoped cuprates is proposed on the 
basis of the $t-t^{\prime}-U$ Hubbard model. The \mbox{prediction} of our theory is consistent 
with the experiment quite well within the \mbox{uncertainty} of the present experimental 
measurement. Therefore we \mbox{argue} that the pseudogap phenomenon in the underdoped 
cuprates can be well explained within the mean field approximation.
\end{abstract}

\mbox{}


\noindent {\bf 1. Introduction}  
\vspace{.5cm}

A large body of experimental investigations have indicated that underdoped 
\mbox{high-temperature} superconductors exhibit intriguing properties at temperatures above the superconducting transition temperature $T_c$. Most notably, the underdoped 
cuprates exhibit a pseudogap behavior below a characteristic temperature $T^\ast$ 
which can be well above the superconducting transition temperature $T_c$. The 
so-called \mbox{``pseudogap''} means a partial gap. ``An example of such a partial 
gap would be a situation where, within the band theory approximation, some regions of 
the Fermi surface \mbox{become} gapped while other parts retain their conducting prope
rties and with \mbox{increased} \mbox{doping} the gapped portion diminishes and the 
materials become more \mbox{metallic}''(quoted from Ref. \cite{1}). What is the origin 
behind it? A number of scenarios like pair \mbox{formation} well above $T_c$ 
\cite{Randeria92a,Emery,Ranninger}, spin--charge separation \cite{Lee,Dai}, 
spin-density wave (SDW) or \mbox{antiferromagnetic} fluctuations \cite{Pines,Kampf} 
have been proposed as possible origins of these pseudogap phenomena. However, no 
consensus has been reached so far, which one is correct in these microscopic theories. 
It should be noted that these theories of the pseudogap are all beyond the mean-field 
approximation.

In present paper, we propose a mean field SDW analysis for understanding the pseudogap phenomena in the underdoped cuprates. Our aim is to examine to what extend can we interpret this phenomenon within mean field theory.

\vspace{.5cm}
\noindent{\bf 2. Electronic band structure} 
\vspace{.5cm}

\begin{figure}[htbp]
\begin{center}
\includegraphics[width=20cm,angle=-90]{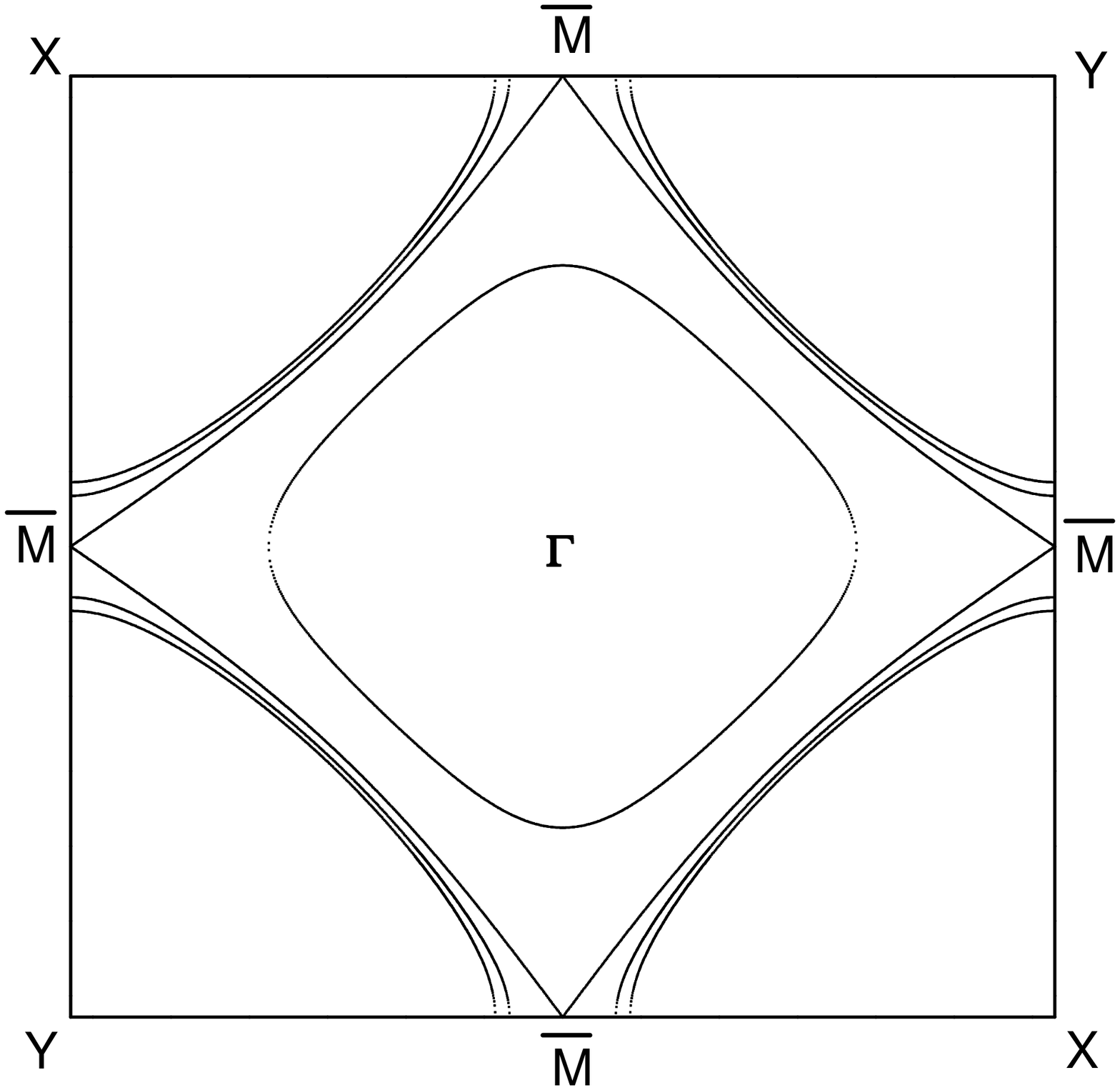}
\caption{}\label{fig1}
\end{center}
\end{figure}

Soon after the discovery of the cuprate superconductors, the electronic band \mbox{structure} of the cuprates has been calculated by the local density approximation (LDA) band calculation \cite{Massidda,Hamada,Freeman,Pickett}. The result of the electronic band structure of the cuprates of the LDA band calculation is consistent with the later angle resolved photoemission experiment \cite{Campuzanoet,Dessau,King,Shen}. The electronic band structure of the cuprates can be well fitted by a tight-binding model, which is written as  
\begin{equation}
\overline{\varepsilon}_{\bf k}=-2t(\cos k_x+\cos k_y)-4t^{\prime}\cos k_x\cos k_y.
\label{1}       
\end{equation}
Where $t$ is nearest-neighbor, $t^\prime$ is next-to-nearest-neighbor. In this paper we consider $t>0$ and $t^{\prime}<0$ only. Energy contour lines for the electronic band structure (\ref{1}) are shown in Fig.$\;$\ref{fig1}. There are two different saddle points locate at the $\overline{M}$ points [($\pm\pi$, 0) and (0, $\pm\pi$)] of the Brillouin zone. The energy contour line with energy $\overline{\varepsilon}_{s}=4t^{\prime}$ pass through the saddle points.

For convenience, we choice $\overline{M}$ as the new origin of the ${\bf k}$-space and take the energy $\overline{\varepsilon}_{s}=4t^{\prime}$ at the saddle point $\overline{M}$ as zero. Then the dispersion (\ref{1}) is reexpressed in the form 
\begin{eqnarray}
\varepsilon_{\bf k}&=&\overline{\varepsilon}_{\bf k}-4t^{\prime}\nonumber\\
&=&-2t(-\cos k_x+\cos k_y)+4t^{\prime}(\cos k_x\cos k_y-1).     
\label{2}       
\end{eqnarray}
If without specific statement, we keep this usage later.

\begin{figure}[htbp]
\begin{center}
\includegraphics[width=20cm,angle=-90]{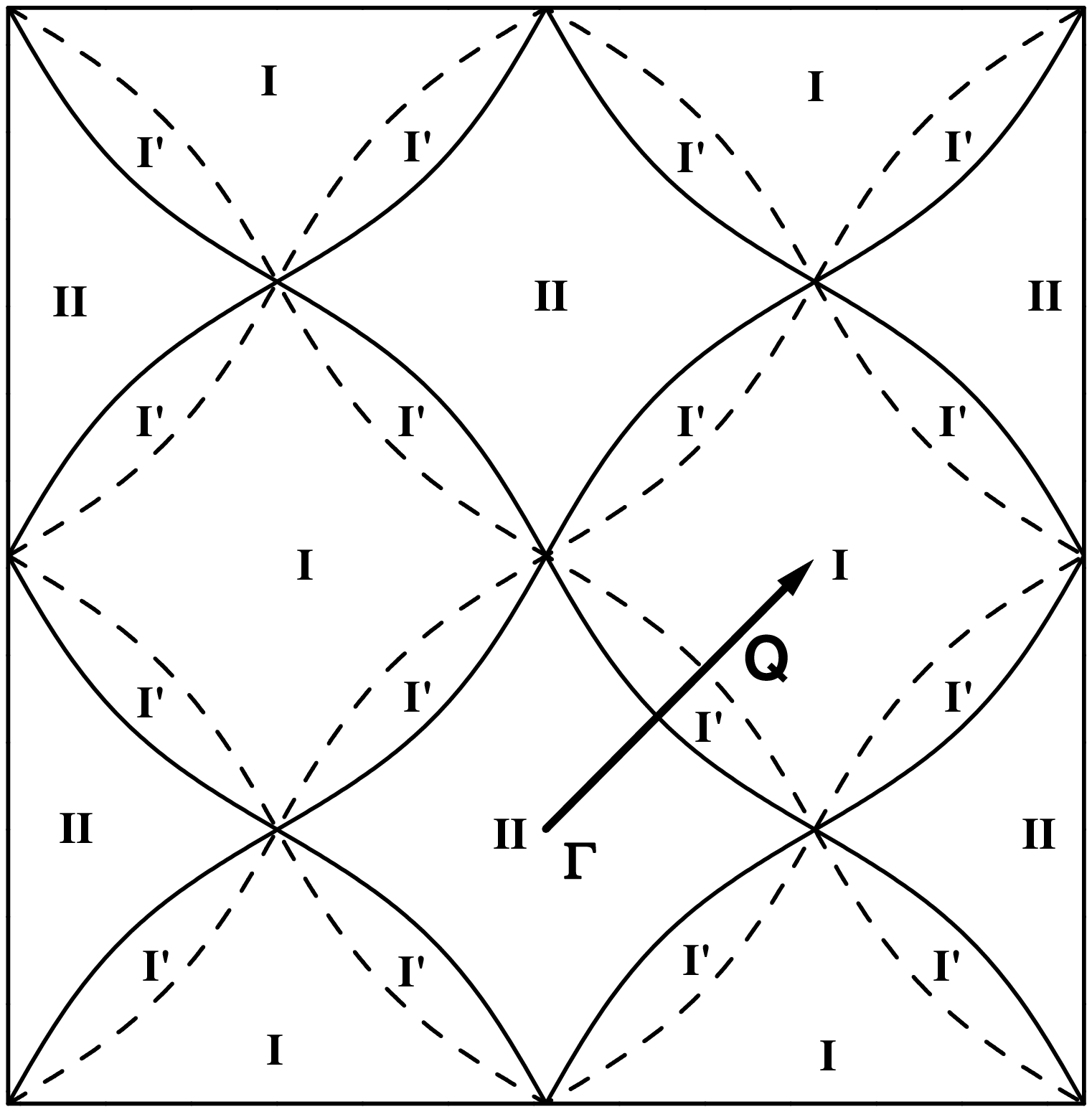}
\caption{}\label{fig2}
\end{center}
\end{figure}

\begin{figure}[htbp]
\begin{center}
\includegraphics[width=20cm,angle=-90]{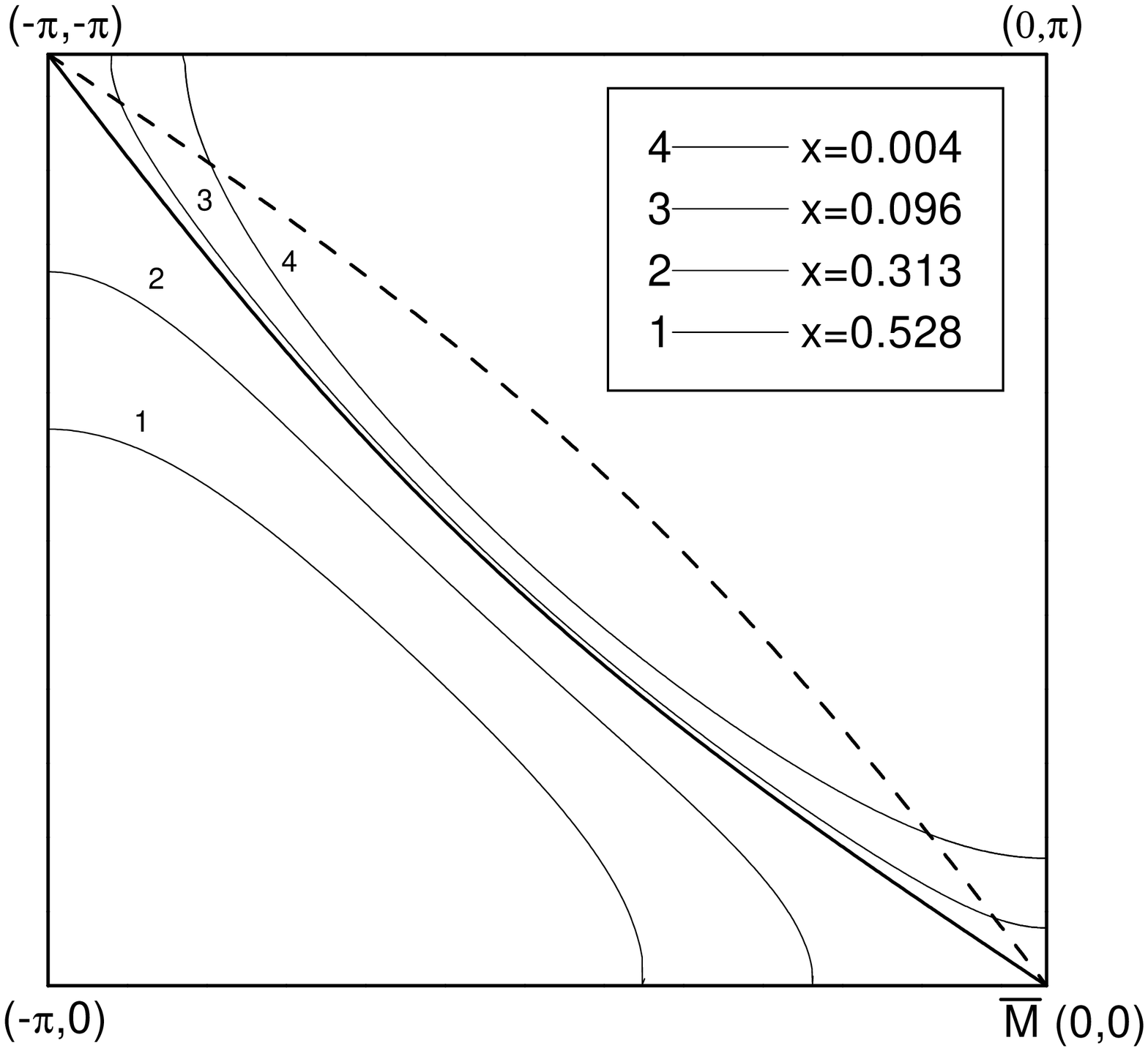}
\caption{}\label{fig3}
\end{center}
\end{figure}

We replot in the period Brillouin zone the energy contour lines passing through the saddle points. As shown in Fig.$\;$\ref{fig2}, there are two different regions: ${\bf I+I}^\prime$ and ${\bf II}$. In the region ${\bf II}$, $\varepsilon_{\bf k}<0$, in the regions ${\bf I+I}^\prime$, $\varepsilon_{\bf k}>0$. The area of the region ${\bf I+I}^\prime$ is larger than that of the region ${\bf II}$. When the region ${\bf II}$ is shifted by the vector ${\bf Q}=(\pi,\pi)$, it coincides with the region ${\bf I}$. The region ${\bf I}'$ is called as the necklace region, which has following features. Firstly, when ${\bf k}$ locates in a bubble, ${\bf k+Q}$ will locate in another one, both $\varepsilon_{\bf k}$ and $\varepsilon_{\bf k+Q}$ are larger then zero. On the other hand, in the regions outside the necklace region ${\bf I}'$, both sign of the $\varepsilon_{\bf k}$ and $\varepsilon_{\bf k+Q}$ are always opposite. For example, when ${\bf k}$ locates in ${\bf I}$, $\varepsilon_{\bf k}>0$, then ${\bf k+Q}$ will locate in ${\bf II}$, $\varepsilon_{\bf k+Q}<0$. Secondly, in the overdoping regime, the Fermi surface entirely lies outside the necklace region (as shown in Fig.$\;$\ref{fig3}). But for the underdoping case, only part of the Fermi surface lies outside the necklace region, and further, with decreased doping the portion outside the necklace region increases (as shown in Fig.$\;$\ref{fig3}).

It is interest to note the fact that when $t^\prime=0$, the necklace region and said \mbox{peculiarity} of the band structure of the cuprates disappears.

\vspace{.5cm}
\noindent{\bf 3. Mean-field theory} 
\vspace{.5cm}

\begin{figure}[htbp]
\begin{center}
\includegraphics[width=20cm,angle=-90]{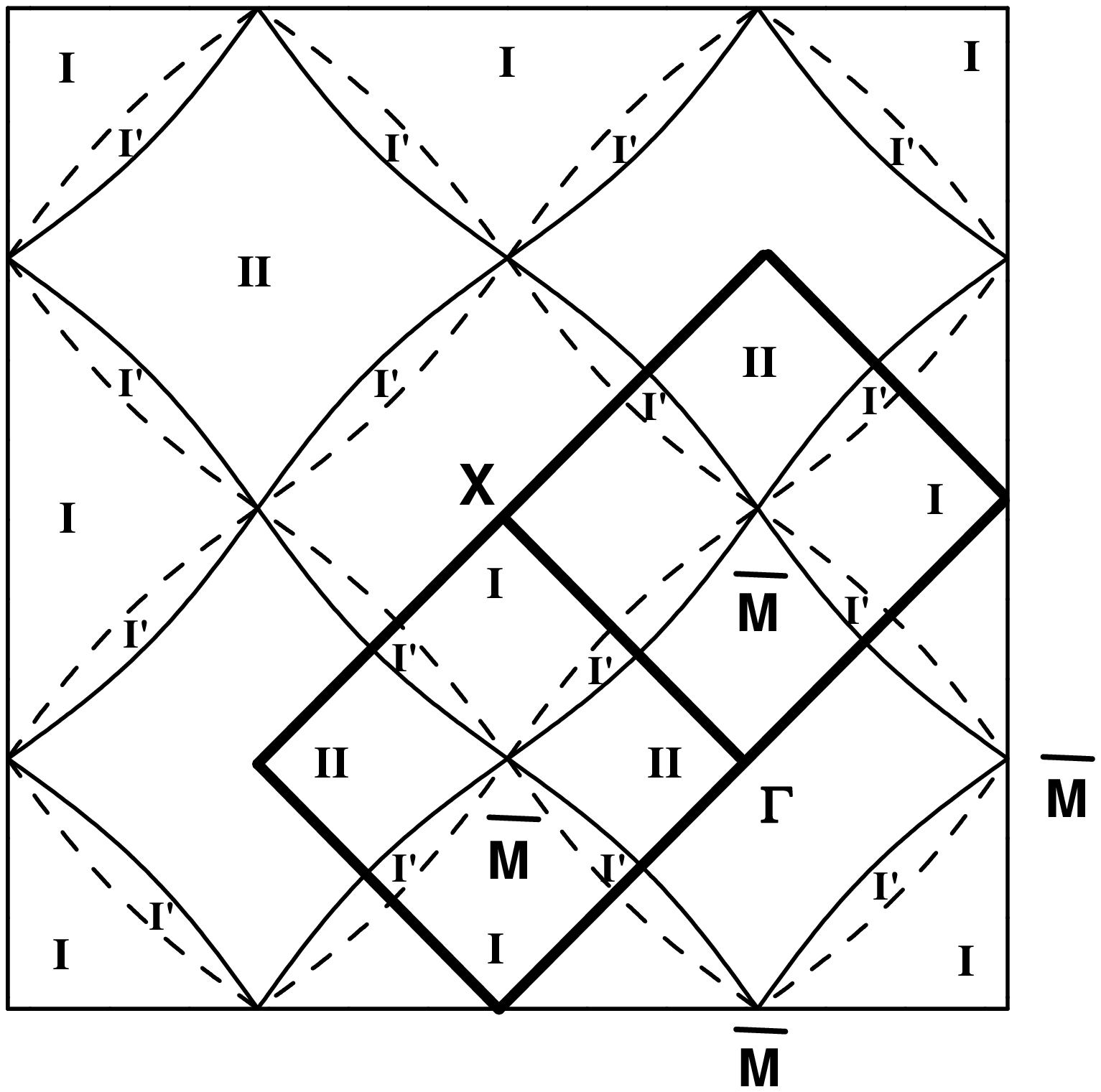}
\caption{}\label{fig4}
\end{center}
\end{figure}

The starting point of our calculation is the Hubbard model. In the momentum representation, the $t-t^{\prime}-U$ Hubbard model can be written as \cite{Fradkin}
\begin{equation}
H=\sum_{{\bf k}\sigma}(\varepsilon_{{\bf k}}-\mu)a_{{\bf k}\sigma}^{\dagger}a_{{\bf k}\sigma}-\;\frac{U}{2N}\sum_{\bf q}\sum_{{\bf k}\sigma{\bf k'}\sigma'}a_{{\bf k+q}\sigma}^{\dagger}\sigma a_{{\bf k}\sigma}a_{{\bf k'}\sigma'}^{\dagger}\sigma' a_{{\bf k'+q}\sigma'}.
\label{3}
\end{equation} 
Here a term $\frac{1}{2}NU$ has been omitted. $U$ is the local Coulomb repulsion. $a_{{\bf k}\sigma}(a_{{\bf k}\sigma}^{\dagger})$ is the annihilation (creation) operator for the electron with momentum ${\bf k}$ and spin $\sigma$. $\mu$ is the chemical potential. $\varepsilon_{{\bf k}}$ is given by Eq.$\;$(\ref{2}). All the momentum summations extend over the Brillouin zone. 
Considering commensurate SDW state and using the mean-field approximation, the Hamiltonian reduces
\begin{equation}
H={\sum_{{\bf k}\sigma}}'(\varepsilon_{{\bf k}}-\mu)a_{{\bf k}\sigma}^{\dagger}a_{{\bf k}\sigma}+{\sum_{{\bf k}\sigma}}'(\varepsilon_{\bf k+Q}-\mu)a_{{\bf k+Q}\sigma}^{\dagger}a_{{\bf k+Q}\sigma}-\Delta {\sum_{{\bf k}\sigma}}'(a_{{\bf k+Q}\sigma}^{\dagger}\sigma a_{{\bf k}\sigma}+h.c).
\label{4}
\end{equation}
Here $\sum_{\bf k}'$ means that the sum extends over the magnetic Brillouin zone (shown in \mbox{Fig.$\;$\ref{fig4}} by the thick square). The term $\frac{N}{2U} \Delta^{2}$ has been omitted.  
The order parameter $\Delta$ is given by
\begin{equation}
\Delta=\frac{2U}{N}{\sum_{{\bf k}\sigma}}'<a_{{\bf k}+{\bf Q}\sigma}^{\dagger}\sigma a_{{\bf k}\sigma}>.
\end{equation}

\begin{figure}[htbp]
\begin{center}
\includegraphics[width=20cm,angle=-90]{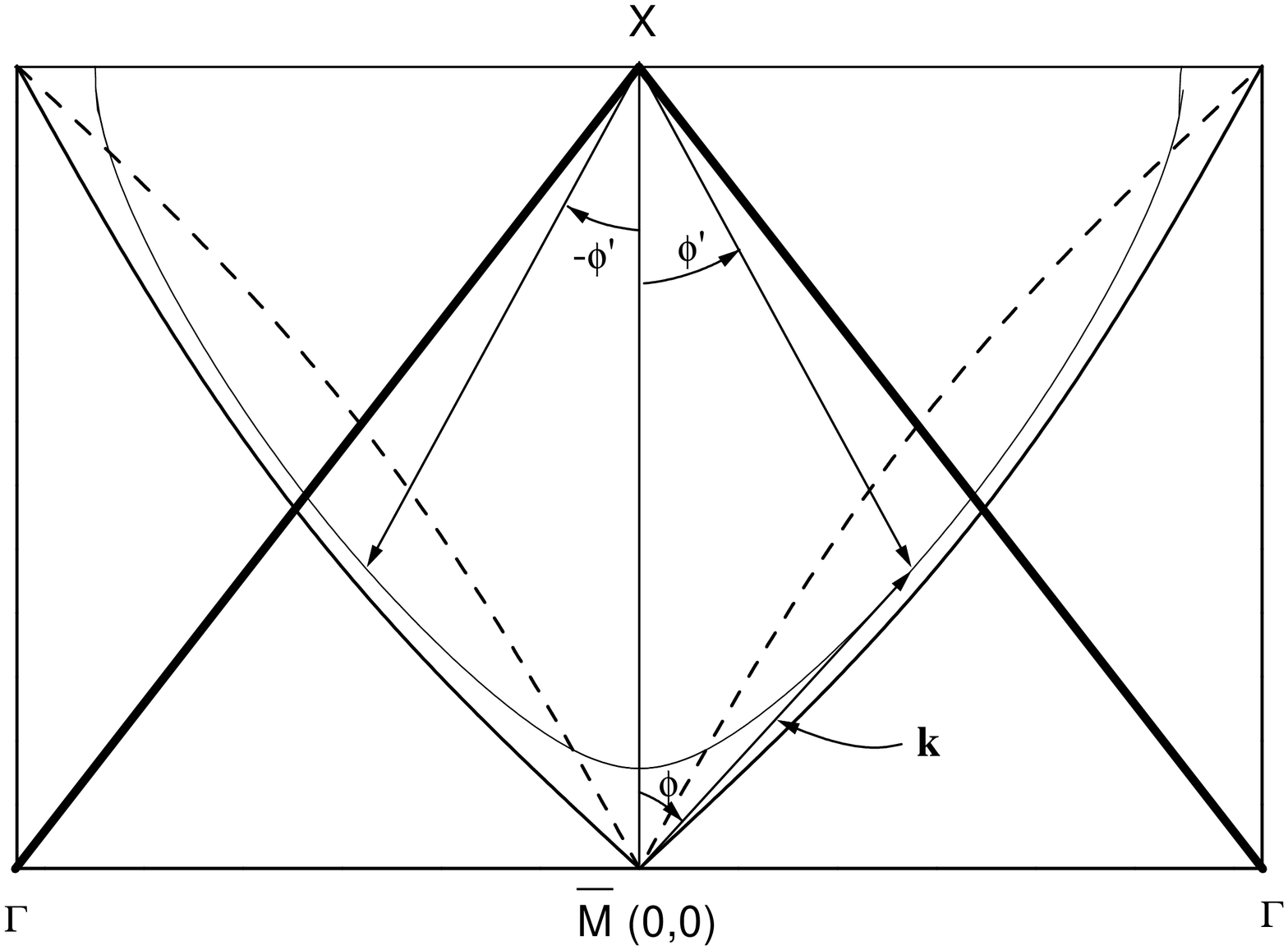}
\caption{}\label{fig5}
\end{center}
\end{figure}

By the following canonical transformation
 
\begin{eqnarray}
\alpha_{\bf k\sigma} & = & u_{\bf k}a_{{\bf k}\sigma}-v_{\bf k}\sigma a_{{\bf k+Q}\sigma}\ , \nonumber \\ 
\gamma_{{\bf k}\sigma}& = & v_{\bf k}\sigma a_{{\bf k}\sigma}+u_{\bf k}\sigma a_{{\bf k+Q}\sigma}\ , 
\label{canon}
\end{eqnarray}
the Hamiltonian (\ref{4}) is diagonalised as 
\begin{equation}
H={\sum_{{\bf k}\sigma}}'(\varepsilon_{1}({\bf k})\alpha^{\dagger}_{{\bf k}\sigma}\alpha_{{\bf k}\sigma}+\varepsilon_{2}({\bf k})\gamma^{\dagger}_{{\bf k}\sigma}\gamma_{{\bf k}\sigma}),
\end{equation}
in which,
\begin{equation}
\varepsilon_{1}({\bf k})=\frac{\varepsilon_{{\bf k}}+\varepsilon_{\bf k+Q}}{2}-\mu+\sqrt{(\frac{\varepsilon_{{\bf k}}-\varepsilon_{\bf k+Q}}{2})^2+\Delta^2},\label{8}
\end{equation}
\begin{equation}
\varepsilon_{2}({\bf k})=\frac{\varepsilon_{{\bf k}}+\varepsilon_{\bf k+Q}}{2}-\mu-\sqrt{(\frac{\varepsilon_{{\bf k}}-\varepsilon_{\bf k+Q}}{2})^2+\Delta^2},\label{9}
\end{equation}
\begin{equation}
\Delta=\frac{U}{N}{\sum_{{\bf k}}}'
\frac{\Delta}{E({\bf k})}
(\tanh (\frac{\varepsilon_{1}({\bf k})}{2T})-
\tanh (\frac{\varepsilon_{2}({\bf k})}{2T}))\label{10}
\end{equation}
and
\begin{equation}
E({\bf k})=\sqrt{(\frac{\varepsilon_{{\bf k}}-\varepsilon_{\bf k+Q}}{2})^2+\Delta^2}.
\label{11}
\end{equation} 
Here $\varepsilon_{1}({\bf k})$ and $\varepsilon_{2}({\bf k})$ are energy dispersions of the quasiparticles. For the hole doping system, the Fermi surface lies inside the lower band ($\varepsilon_{2}({\bf k})$). The pseudogap is given by 
 
\begin{eqnarray} 
\Delta_{PS}(\phi)&=&|\varepsilon_{2}({\bf k})|\nonumber\\
&=&
\mu-4t^{\prime}(\cos k_x\cos k_y-1)+\sqrt{4t^2 (\cos k_x-\cos k_y)^2+\Delta^2}. 
\label{12}
\end{eqnarray}
In Fig.$\;$\ref{fig5} we plot the part of the magnetic Brillouin zone of the Fig.$\;$\ref{fig4}. The light curve represents the Fermi surface. $k_x$- and $k_y$-axis are parallel with $\overline{M}\Gamma$ and $\overline{M}$X, respectively. In Eq.$\;$(\ref{12}), ${\bf k}=(k_x, k_y)$ is the wave vector of the Fermi surface, i.e. $\varepsilon_{{\bf k}}-\mu=0$. $\phi=\arctan (k_x/k_y)$ is polar angle of the wave vector ${\bf k}$. For convenience, we take $\phi^{\prime}=\arctan(k_x/(\pi-k_y))$ as variable instead of the $\phi$ in the following calculations.

\vspace{.5cm}
\noindent {\bf 4. Results} 
\vspace{.5cm}

In this section, we analyse the angular dependence of the pseudogap $\Delta_{PS}(\phi')$ along the Fermi surface. Ouing to the symmetry of the energy spectrum $\varepsilon_{2}({\bf k})$, our analysis can be limited only in the interval $0\leq\phi'\leq45^\circ$.

\begin{figure}[htbp]
\begin{center}
\includegraphics[width=20cm,angle=-90]{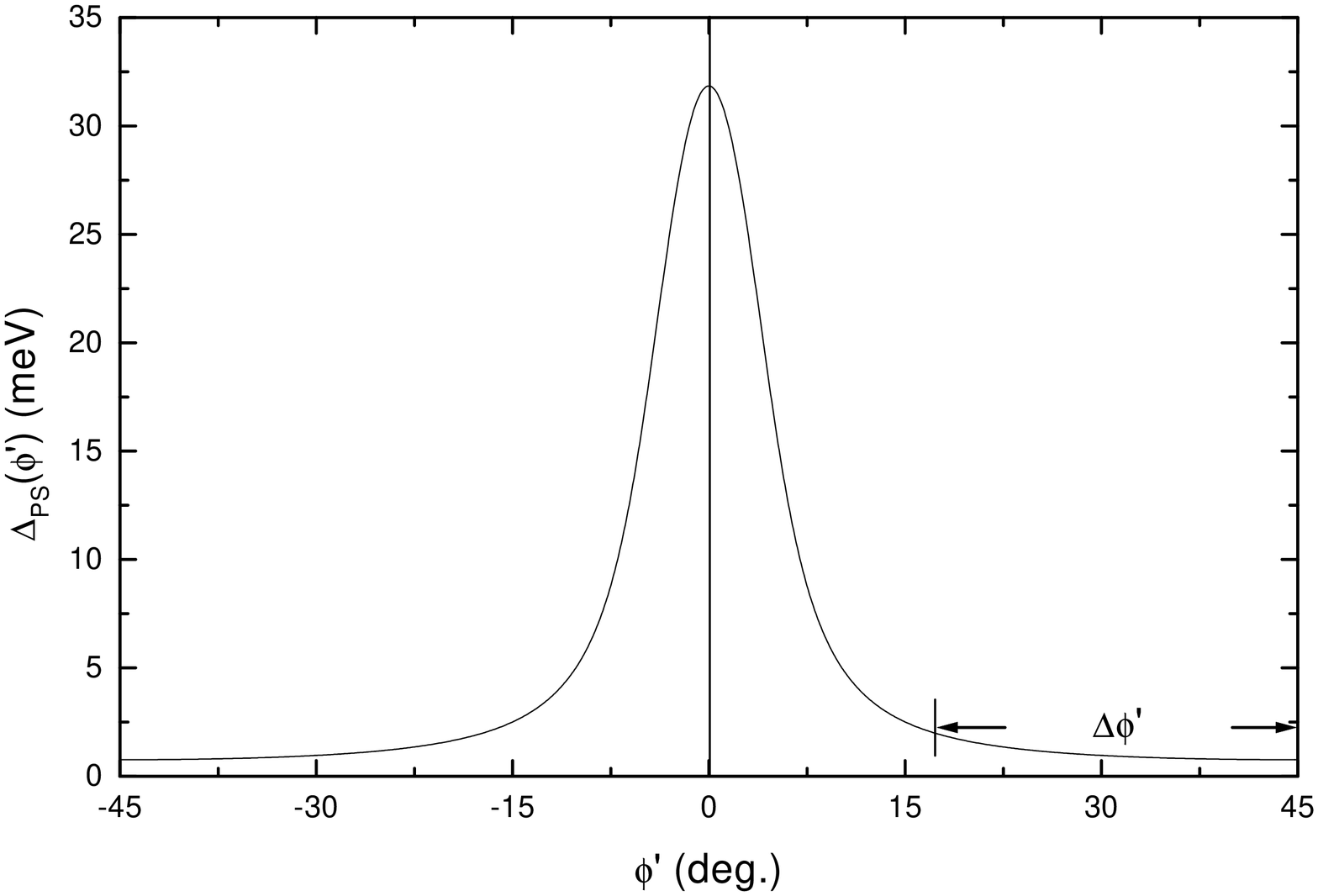}
\caption{}\label{fig6}
\end{center}
\end{figure}

By solving Eqs.$\;$(\ref{8}), (\ref{9}), (\ref{10}), (\ref{11}) and (\ref{12}) numerically, we compute $\Delta_{PS}(\phi')$ at $T=0$ K in the underdoping regime ($\mu>0$). In the computation, we choose \mbox{$t=430$ meV}, $t'/t=-0.18$, $U/t=0.8$ and the hole doping concentration x=0.13. The results are plotted as $\Delta_{PS}(\phi')$ versus $\phi^{\prime}$ curve in Fig.$\;$\ref{fig6}. It shows that there is strong angular dependence of the $\Delta_{PS}(\phi')$, as one moves along the Fermi surface from $\phi^{\prime}=0$ (i.e. near the saddle point $\overline{M}$, or at the hot spot) to $\phi^{\prime}=45^\circ$ (i.e. cold spot). At first, we see the maximum pseudogap at $\phi^{\prime}=0$. As we go into the necklace region, the pseudogap drops quickly and, at approximately $18^\circ$, drops down to 2 meV. And then, the pseudogap decreases monotonously to $\Delta_{PS}(45^\circ)$. Experimental measurement \mbox{reveals} that only a portion of the Fermi surface near the saddle point $\overline{M}$ becomes gapped while in other parts, the pseudogap is equal to zero \cite{1,Norman}. However, the \mbox{error-bar} of the pseudogap data is rather larger\footnote{$^)$See, for example, Fig.$\;$\ref{fig8} of paper \cite{1}}$^)$. It is impossible to say certainly that along the part of the Fermi surface near the cold point, the pseudogap is real zero or only a small quantity. Keeping this fact in mind, we conclude that the general structure of the pseudogap along the Fermi surface, shown in Fig.$\;$\ref{fig6}, captures the main feature of experiment \cite{Norman}\footnote{$^)$Ref. also the review article \cite{1} and the papers listing in it}$^)$.

\begin{figure}[htbp]
\begin{center}
\includegraphics[width=20cm,angle=-90]{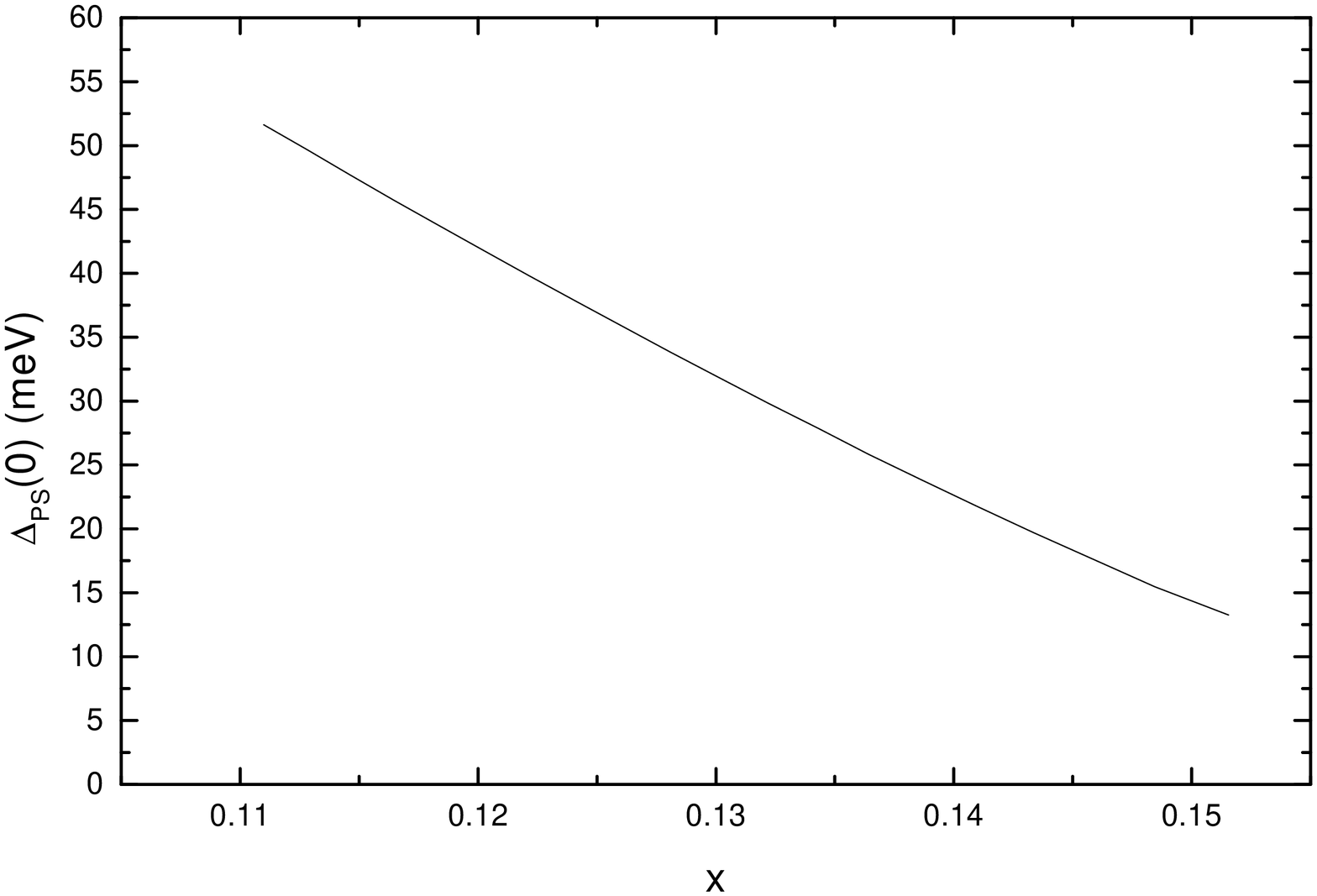}
\caption{}\label{fig7}
\end{center}
\end{figure}

\begin{figure}[htbp]
\begin{center}
\includegraphics[width=20cm,angle=-90]{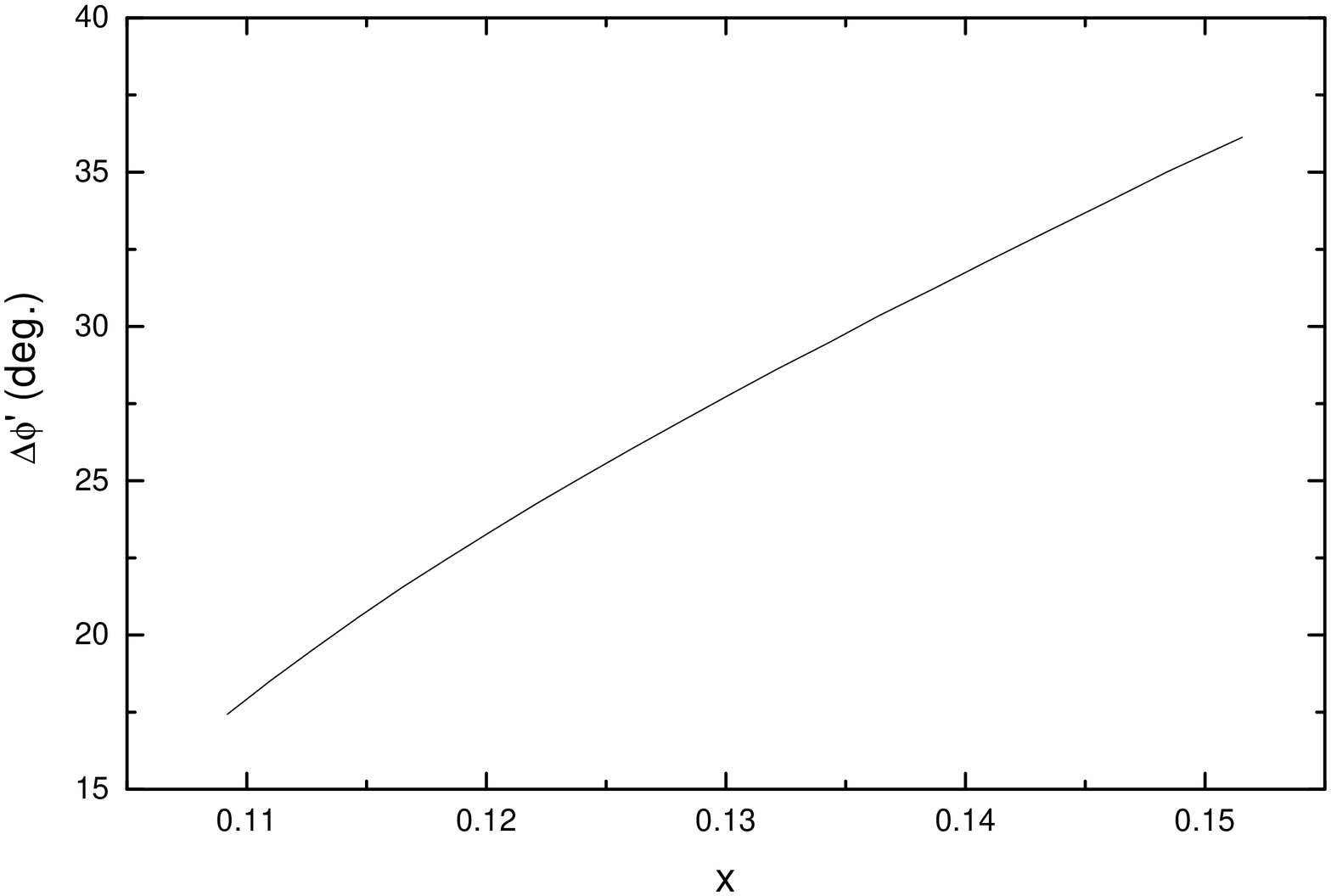}
\caption{}\label{fig8}
\end{center}
\end{figure}

The dependence of $\Delta_{PS}(0)$ on the hole doping concentration is shown in Fig.$\;$\ref{fig7}. It shows that $\Delta_{PS}(0)$ increase with the decrease of hole doping.
In Fig.$\;$\ref{fig8}, we plot $\Delta\phi'$ versus the hole doping concentration curve. Here, $\Delta\phi'$ is the interval $\phi^{\prime}$ (measured from $\phi^{\prime}=45^\circ$), defined by the requirement that the value $\Delta_{PS}(\phi')$ is less than a proper chosen value (say, 2 meV in Fig.$\;$\ref{fig6} and \ref{fig8}). It can be seen from Fig.$\;$\ref{fig8} that the length of the Fermi arc, along which the pseudogap less than 2 meV, increase with the increase of doping. It implies that as doping increase, the portion of the Fermi surface destroyed by the pseudogap decreases. The prediction discribed above is consistent with the experiment \cite{Norman}$^{2)}$.

\vspace{.5cm}
\noindent {\bf 5. Concluding remarks} 
\vspace{.5cm}

It is of interest to note that the situation is entirely different if $t^{\prime}=0$. For in this case, Eq.$\;$(\ref{12}) reduces to 
\begin{equation}
\Delta_{PS}(\phi^{\prime})=\mu+\sqrt{\mu^{2}+\Delta^{2}}.\label{13}
\end{equation}
It is in contradiction with the experiment \cite{1,Norman}, for the pseudogap along the Fermi surface, according to (\ref{13}), is constant.

Now, it is clearly that the peculiarity of the band structure of the cuprate plays an important role in understanding the pseudogap phenomenon in underdoped cuprate. This is the reason why our mean field SDW analysis of the \mbox{pseudogap}, based on the $t-t^{\prime}-U$ Hubbard model, meets with success.

The mean-field solution has an antiferromagnetic long-range order. At sufficient doping concentration, the spin long-range order will be removed by fluctuations but there are still short-range orderings. We assume implicitly in our theory that the \mbox{pseudogap} structure, at least near the saddle point ($\pi,0$), is not sensitive to the \mbox{long-range} order and will survive in underdoped region, leading to the pseudogap \mbox{phenomenon}.

\vspace{2cm}
\begin {center}
{\large{\bf Acknowledgments}}
\end {center}

One of the authors (H. S. Wu) would like to thank Prof. Z. Y. Weng for very valuable discussion.

\newpage

\begin{center}
{\bf Figure Captions:}
\end{center}

\begin{description}
\item{Fig.$\,$\ref{fig1}.} The Brillouin zone and energy contour lines: The $\Gamma$ point is at the middle of the Brillouin zone and the $\overline{M}$ points [($\pm\pi$, 0) and (0, $\pm\pi$)] are midway along the edges. The curves are the energy contour lines ($t^{\prime}/t=-0.16$) with energy \mbox{$\overline{\varepsilon}_{\bf k}/t=-1.59, -0.64, -0.49$ and $-0.4$}, which are from inside to outside.
 
\item{Fig.$\;$\ref{fig2}.} The period Brillouin zone: The solid curves are the energy contour lines with $\varepsilon_{\bf k}=0$. When the region ${\bf II}$ is shifted by the vector ${\bf Q}=(\pi,\pi)$, it coincides with the region ${\bf I}$. In regions ${\bf I+I'}$, $\varepsilon_{\bf k}>0$. In the region ${\bf II}$, $\varepsilon_{\bf k}<0$. The region ${\bf I}'$ is called as the necklace region.

\item{Fig.$\;$\ref{fig3}.} The Fermi surfaces ($t^{\prime}/t=-0.16$) in the quarter of Brillouin zone: The light curve represents the Fermi surface. 1 and 2 for the overdoping. 3 and 4 for the underdoping. The heavy dashed and the solid curve represent the necklace region boundary.
 
\item{Fig.$\;$\ref{fig4}.} Our choice of the Brillouin zone. The heavy rectangle is our choice of the Brillouin zone boundary and the origin is at the $\overline{M}$. The heavy square is the magnetic Brillouin zone boundary.
 
\item{Fig.$\;$\ref{fig5}.} This figure is the part of the magnetic Brillouin zone of the Fig. 4. The light curve represents the Fermi surface for the underdoping region ($t^{\prime}/t=-0.16$). The dashed and the solid curve represent the necklace region boundary. The heavy solid lines are the magnetic Brillouin zone boundary. $k_x$- and $k_y$-axis are parallel with $\overline{M}\Gamma$ and $\overline{M}$X, respectively. $\phi=\arctan (k_x/k_y)$. $\phi^{\prime}=\arctan(k_x/(\pi-k_y))$.
  
\item{Fig.$\;$\ref{fig6}.} The angle dependence of the pseudogap located at $\phi^{\prime}$, $\Delta_{PS}(\phi')$, for the hole doping concentration x=0.13 ($t^{\prime}/t=-0.18$ and $U/t=0.8$). The $\Delta\phi'$ is the region where the values of $\Delta_{PS}(\phi')$ are all smaller than 2 meV.
 
\item{Fig.$\;$\ref{fig7}.} The hole doping concentration dependence of the pseudogap located at \mbox{$\phi^{\prime}=0$}, $\Delta_{PS}(0)$, for $t^{\prime}/t=-0.18$ and $U/t=0.8$. The x indicates the hole doping \mbox{concentration}. The solid curve represents the pseudogap in the underdoping region.
 
\item{Fig.$\;$\ref{fig8}.} The hole doping concentration dependence of the region where the values of $\Delta_{PS}(\phi')$ are all smaller than 2 meV, $\Delta\phi'$, for the underdoping region ($t^{\prime}/t=-0.18$ and $U/t=0.8$). The x indicates the hole doping concentration.
\end{description}

\newpage

\end{document}